\def\jnl#1#2#3#4{#1 {\bf #2}, #3 (#4).}

\def\JPCS{J.\ Phys.\ Chem.\ Solids}
\def\JPSJ{J.\ Phys.\ Soc.\ Jpn.}

\def\PRB{Phys.\ Rev.\ B}
\def\PRL{Phys.\ Rev.\ Lett.}

\def\beq{\begin{equation}}
\def\eeq{\end{equation}}
\def\beqa{\begin{eqnarray}}
\def\eeqa{\end{eqnarray}}
\def\subbeqa{\begin{subeqnarray}}
\def\subeeqa{\end{subeqnarray}}

\newcommand{\btwo}{Bi$_2$Sr$_2$CaCu$_2$O$_{\rm 8+y}$}

\newcommand{\bfou}{(Cu$_{0.6}$C$_{0.4}$)Bi$_2$Sr$_2$Ca$_3$Cu$_4$O$_{\rm 12+y}$}

\newcommand{\cuo}{CuO$_2$ }

\newcommand{\etal}{{\it et al.} }
\newcommand{\ie}{{\rm i.e.}}

\newcommand{\tc}{{\it T$_c$}}
\newcommand{\tj}{{\it t-t$'$-t$''$-J} }
\newcommand{\nip}{$N_h$(IP)}
\newcommand{\nop}{$N_h$(OP)}
\newcommand{\ndif}{$\Delta N_h\,$}
\newcommand{\hnu}{$h\nu$ }
\documentstyle[aps,prb,multicol,floats,epsf]{revtex}

\begin{document}
\draft
\wideabs{
\title{
Electronic states and superconductivity in multilayer high-$T_c$ cuprates
}
\author{
M. Mori, T. Tohyama, and S. Maekawa
}
\address{
Institute for Materials Research, Tohoku University, Sendai 980-8577, Japan
}

\date{\today}
\maketitle
\begin{abstract}
We study electronic states of multilayer cuprates in the normal phases as functions of the number of \cuo planes and the doping rate. 
The resonating valence bond wave function and the Gutzwiller approximation are used for a two-dimensional multilayer \tj model. 
We calculate the electron-removal spectral functions at ($\pi$,0) in the \cuo plane next to the surface to understand the angle-resolved photoemission spectroscopy (ARPES) spectra. 
We find that the trilayer spectrum is narrower than the bilayer spectrum but is wider than the monolayer spectrum.
In the tri- and tetralayer systems, the outer \cuo plane has different superconducting amplitude from the inner \cuo plane, while each layer in the bilayer systems has same amplitude. 
The recent ARPES and NMR experiments are discussed in the light of the present theory.
\end{abstract}
\pacs{PACS number: 74.25.Jb, 74.20.-z, 71.10.Fd.
\hspace{72pt} \underline{Phys. Rev. B {\bf 66}, 064502 (2002)}.}
}
\narrowtext
\section{Introduction}\label{Intro}
In studies of high-\tc$\;$ superconductivity, the number of \cuo planes in a unit cell, $n$, is one of the key factors. 
It is indicated that the maximum \tc$\;$ monotonically increases with $n$ and saturates at either $n$=3 or 4.\cite{Ihara,Parkin} 
The bilayer compounds have crystallographically equivalent \cuo planes, while the tri- and tetralayer compounds have inequivalent planes, \ie, the outer pyramidal coordinated planes (OP) and the inner square coordinated planes (IP). 
By calculating the Madelung energy for Bi and Tl compounds in the point-charge model, it was shown that the hole density of the IP, \nip, is lower than that of the OP, \nop, since the IP is sandwiched by Ca sheets with positive 
charge.\cite{Kondo,Ohta2}
It was also argued that a hole density difference $\Delta N_h \equiv N_h({\rm OP})-N_h({\rm IP})$ becomes large by increasing $n$ and controls the superconducting phases.\cite{Ohta,DiStatio,Haines}

Such hole distribution among the \cuo layers was observed above \tc$\,$ by the $^{17}$O and $^{63}$Cu nuclear magnetic resonance (NMR) measurement for Pb-doped 
Bi$_2$Sr$_2$Ca$_2$Cu$_3$O$_{\rm 10+y}$ (Bi2223),\cite{Trokiner,Statt} and for 
Hg Sr$_2$Ca$_2$Cu$_3$O$_{\rm 10+y}$ (Hg1223).\cite{Magishi,Julien} These studies showed that the IP has smaller carrier density and larger antiferromagnetic (AF) correlation than the OP.
Kotegawa \etal systematically estimated the hole density in each layer by the $^{63}$Cu Knight shift in HgBa$_2$Ca$_{\rm n-1}$Cu$_{\rm n}$O$_{\rm 2n+2+y}$ and CuBa$_2$Ca$_{\rm n-1}$Cu$_{\rm n}$O$_{\rm 2n+4-y}$.\cite{Kotegawa}
By use of an empirical relation between the Knight shift at room temperature and the hole density, which was deduced from the nuclear quadruple resonance (NQR) frequency,\cite{Zheng} they found that the hole density difference among the layers increases as either a total carrier content $\delta_{total}$ or $n$ increases.

As regards an interlayer hopping, the bilayer splitting was shown by band calculations\cite{Chakravarty,OKAnderson,Liechtenstein} and clearly observed by the angle-resolved photoemission spectroscopy (ARPES) in \btwo (Bi2212) above \tc.\cite{Feng1,Chuang1,Kordyuk}
However, a trilayer splitting  in Bi2223 has not been observed.\cite{Feng2,Sato} To understand the electronic states in the multilayer cuprates, it is necessary to clarify the reason why the trilayer splitting is not clearly observed, although the bilayer splitting is distinguishable. 

In the present paper,  
the electron-removal spectral function is calculated in a two-dimensional multilayer \tj model at zero temperature by using the resonating valence bond (RVB) wave function and the Gutzwiller approximation.
We study the $n$ dependence of the single-particle excitations including the interlayer hopping which connects the chemically different \cuo planes for $n\ge3$. Since the ARPES measurement is sensitive to a surface of sample due to a short escape depth of outgoing electron, the spectra in the tri- and tetralayer compounds should be dominated by the OP's contribution. 
Moreover, the spectra depend on the incident photon energy \hnu and the photoemission matrix element.\cite{Chuang2,Feng3,Dahnken,Lindroos} 
The spectra in the bilayer systems have two peaks composed of the bonding (B) and anti-bonding (AB) bands, while the spectra in trilayer systems have three peaks. 
We find that the splitting between the two dominant peaks in the trilayer system is smaller than the bilayer splitting. Then, the trilayer spectrum is narrower than the bilayer spectrum but is wider than the monolayer spectrum. 

In the superconducting (SC) phase, Tokunaga \etal studied the temperature dependence of the Knight shift ($K_s$) and the nuclear spin-lattice relaxation rate (1/$T_1$) of $^{63}$Cu in the tetralayer compound, \bfou,\cite{Tokunaga} and found two characteristic temperatures associated with the IP and the OP, $T_0$(IP) and $T_0$(OP). Both $K_s$ and 1/$T_1$ in the IP markedly decrease below $T_0({\rm IP})$=117 K, while those in the OP moderately decrease below $T_0({\rm IP})$=117 K, but markedly decrease below $T_0({\rm OP})$=60 K. These results suggest that a bulk SC transition set in at 117 K but the temperature dependence of SC amplitude in the OP is different from the BCS theory. Such a non-BCS amplitude is found in a two-band system where the SC amplitude in one band is different from that in the other band, and these bands are coupled to each other by a pair-transfer interaction.\cite{Suhl} The two kinds of $T_0$'s are observed in the tetralayer compounds, while it is not found in the trilayer compounds.\cite{Kotegawa} We calculate the SC amplitude of each layer and find that the outer \cuo plane in both the tri- and tetralayer systems has different SC amplitude from the inner \cuo plane. In our calculations, the difference of SC amplitude in the trilayer systems is comparable to that in the tetralayer systems. We consider that the $T_0({\rm IP})$ different from $T_0({\rm OP})$ could be observed also in the trilayer compounds.

The rest of this paper is organized as follows.
In Sec. \ref{model}, we introduce the multilayer \tj model that includes the interlayer hopping and the site potential. 
Section \ref{spectrum} is devoted to show the electronic states in the normal phases and the electron-removal spectral functions. 
The layer dependence of superconducting amplitudes is shown in Sec. \ref{super}.
Summary and discussions are given in Sec. \ref{summary}.
\section{Theoretical Model}\label{model}
We examine the electronic state in the characteristic block of $n$ \cuo layers. 
The Hamiltonian of each layer is given by 
\beqa
\hat{H}_{intra}
	&=&
	-t\sum_{\alpha,\langle ij\rangle_{1st},\sigma}
		\hat{c}^{\dag}_{\alpha,j,\sigma}\hat{c}_{\alpha,i,\sigma}
	-t'\sum_{\alpha,\langle ij\rangle_{2nd},\sigma}
		\hat{c}^{\dag}_{\alpha,j,\sigma}\hat{c}_{\alpha,i,\sigma}\nonumber\\
&&	-t''\sum_{\alpha,\langle ij\rangle_{3rd},\sigma}
		\hat{c}^{\dag}_{\alpha,j,\sigma}\hat{c}_{\alpha,i,\sigma} + {\rm H.c.}\nonumber\\
&& +   J\sum_{\alpha,\langle ij\rangle_{1st}}
		\vec{S}_{\alpha,i} \cdot \vec{S}_{\alpha,j}, \label{intra}
\eeqa
where $\hat{c}_{\alpha,i,\sigma}=c_{\alpha,i,\sigma}(1-n_{\alpha,i,-\sigma})\,$ is the annihilation operator of an electron in the $\alpha$ layer with spin $\sigma$ at site $i$ with the constraint of no double occupancy, and $n_{\alpha,i,-\sigma}$ and $\vec{S}_{\alpha,i}$ are the charge and the spin operators, respectively. 
The summations $\langle ij\rangle_{1st}$, $\langle ij\rangle_{2nd}$, and $\langle ij\rangle_{3rd}$ run over first, second, and third nearest-neighbor sites, respectively. 
The values of the parameters are as follows; $J=0.14$ eV, $t/J=2.5$, $t'/J=-0.85$, and $t''/J=0.575$.\cite{Kim,Tohyama1}

The interlayer hopping has the dispersion relation\cite{Chakravarty}
\beq
\epsilon_{\perp}(k)
	=
	-\frac{t_{\perp}}{4}
	\left[
	\cos(k_x)-\cos(k_y)
	\right]^2,\label{interhop}
\eeq
which is obtained by integrating out the high-energy degrees of freedom from the  Hamiltonian reproducing the full local-density approximation (LDA) calculation.\cite{OKAnderson,Liechtenstein} 
We adopt Eq.~(\ref{interhop}) for $n\ge 2$ and $t_{\perp}/J=1.0$,\cite{Chakravarty,OKAnderson,Liechtenstein,Feng1,Chuang1} and write the Hamiltonian of the interlayer hopping as
\beq
\hat{H}_{inter}
	=
	\sum_{\alpha\neq\beta,k,\sigma} \epsilon_{\perp}(k)
	\left(\hat{c}^{\dag}_{\alpha,k,\sigma}\hat{c}_{\beta,k,\sigma} + {\rm H.c.}\right),\label{inter}
\eeq
where $\alpha$ and $\beta$ are the neighboring layer indices. 

To describe the hole distribution among the layers, we introduce a site potential $W$ in the IP's as
\beq
\hat{H}_{site}
	=
	W \sum_{\gamma={\rm in},k,\sigma}
		\hat{c}^{\dag}_{\gamma,k,\sigma}\hat{c}_{\gamma,k,\sigma}
	=
	W \sum_{\gamma={\rm in}}n_{\gamma}, \label{site}
\eeq
where $\gamma$ indicates the IP index. 
The value of $W$ should be negative. 
For simplicity, $W$ is taken to be independent of the hole densities.

The total Hamiltonian is
\beq
\hat{H} =\hat{H}_{intra}+\hat{H}_{inter}+\hat{H}_{site}.\label{Htot}
\eeq
Since this Hamiltonian includes the constraint of no double occupancy, we adopt the Gutzwiller approximation and statistically average the constraint before variational calculation. Details of calculations are summarized in the Appendix.
In the normal phases, the order parameter, 
\beq
\chi_{\alpha,\tau}
	\equiv
	\frac{1}{N}\sum_i
	\langle 
	c^{\dag}_{\alpha,i,\uparrow  }c_{\alpha,i+\tau,\uparrow}
	+c^{\dag}_{\alpha,i,\downarrow}c_{\alpha,i+\tau,\downarrow}
	\rangle, \label{uRVB}
\eeq 
is introduced and referred to as $\lq\lq$uniform RVB$"$ (URVB).  
In addition to this, in the SC phase the order parameter,
\beq
B_{\alpha,\tau}
	=
		-\frac{1}{N}\sum_i
		\langle
		 c_{\alpha,i,\uparrow  }c_{\alpha,i+\tau,\downarrow}
		-c_{\alpha,i,\downarrow     }c_{\alpha,i+\tau,\uparrow  }
		\rangle, \label{sRVB}
\eeq
is included and referred to as $\lq\lq$singlet RVB$"$ (SRVB). 

Then, we obtain the following mean-field Hamiltonian, which is numerically solved: 
\beqa
\hat{H}_{RVB}&-&\mu N\nonumber\\
 	&=&
	\sum_{\alpha,k,\sigma}	 \xi_{\alpha}(k)
		c^{\dag}_{\alpha,k,\sigma}c_{\alpha,k,\sigma}\nonumber\\
	&+&
	\sum_{\alpha\ne\beta,k,\sigma}	 \epsilon_{\perp}(k)
		c^{\dag}_{\alpha,k,\sigma}c_{\beta,k,\sigma}
	+
	\sum_{\gamma,k,\sigma} W 
		c^{\dag}_{\gamma,k,\sigma}c_{\gamma,k,\sigma}\nonumber\\
	&-&\sum_{k,\tau=x,y}\frac{3}{4}\hat{J}
 		\Big[
 		B^*_{\alpha,\tau} \cos(k_{\tau})
 		 c_{\alpha,-k,\downarrow}c_{\alpha,k,\uparrow}\nonumber\\
 	&&\hspace{42pt}+ 
 		B  _{\alpha,\tau} \cos(k_{\tau})
 		 c^{\dag}_{\alpha,k,\uparrow  }c^{\dag}_{\alpha,-k,\downarrow}
		\Big],\\
\xi_{\alpha}(k)
	&=&
	-\Big\{2\hat{t}  \left[\cos(k_x)+\cos(k_y)\right]+4\hat{t}' \cos(k_x)\cos(k_y)\nonumber\\
  	&&
  	+2\hat{t}''\left[\cos(2k_x)+\cos(2k_y)\right]\Big\} - \mu\nonumber\\
	&&
	-\frac{3}{4}\hat{J}
	\left[\chi_{\alpha,x}\cos(k_x)+\chi_{\alpha,y}\cos(k_y)\right],\\
\chi_{\alpha,\tau}
	&=&
		\frac{1}{N}\sum_k \sum_{\sigma} \cos(k_{\tau})
			\langle c^{\dag}_{\alpha,k,\sigma}c_{\alpha,k,\sigma} \rangle,\\
B_{\alpha,\tau}
	&=&
		-\frac{1}{N}\sum_k 2\cos(k_{\tau})
		\langle
		 c_{\alpha,k,\uparrow  }c_{\alpha,-k,\downarrow     }
		\rangle,
\eeqa
where $\mu$ is the chemical potential and $N$ is the total electron number. 
The values of hopping integrals and the magnetic coupling constant are renormalized by the Gutzwiller factors as
\beqa
&&
\hat{t}= g_t\cdot t, \;
\hat{t}'= g_t\cdot t', \;
\hat{t}''= g_t \cdot t'', \;
\hat{t}_{\perp}= g_t \cdot t_{\perp},\nonumber\\
&&
\hat{J}= g_J \cdot J,
\eeqa
and
\beq
g_t = 2\delta/(1+\delta), \;g_J=4/(1+\delta)^2.
\eeq
The even parity is imposed on $\chi_{\alpha,\tau}$. 
The $k$ summations are carried out on 160$\times$160 points in the first Brillouin zone.

\section{Electronic States in Normal Phases}\label{spectrum}
We study single-particle excitations to clarify the electronic states in the normal phases. 
The characteristics of multilayer systems are manifest at ($\pi$,0) in the dispersion relations, since the interlayer splitting is largest at this point according to Eq.~(\ref{interhop}). 
The electron-removal spectral functions at ($\pi$,0) are shown in this section.
The averaged doping rate $\delta$ is defined as $\delta=\delta_{total}/n$, in which $\delta_{total}$ is a total hole density in the characteristic block of $n$ \cuo layers. 
We choose $\delta=0.25$ or 0.3, below. 
These values of $\delta$ are useful to study the $n$ dependence of the single-particle excitations, since the ARPES spectra at ($\pi$,0) show sharp quasiparticle peaks for a large doping rate. 
On the other hand, the ARPES spectra become broad and dull by decreasing the doping rate.\cite{Shen,Campuzano} 

\begin{figure}
\epsfxsize=7.5cm
\centerline{\epsffile{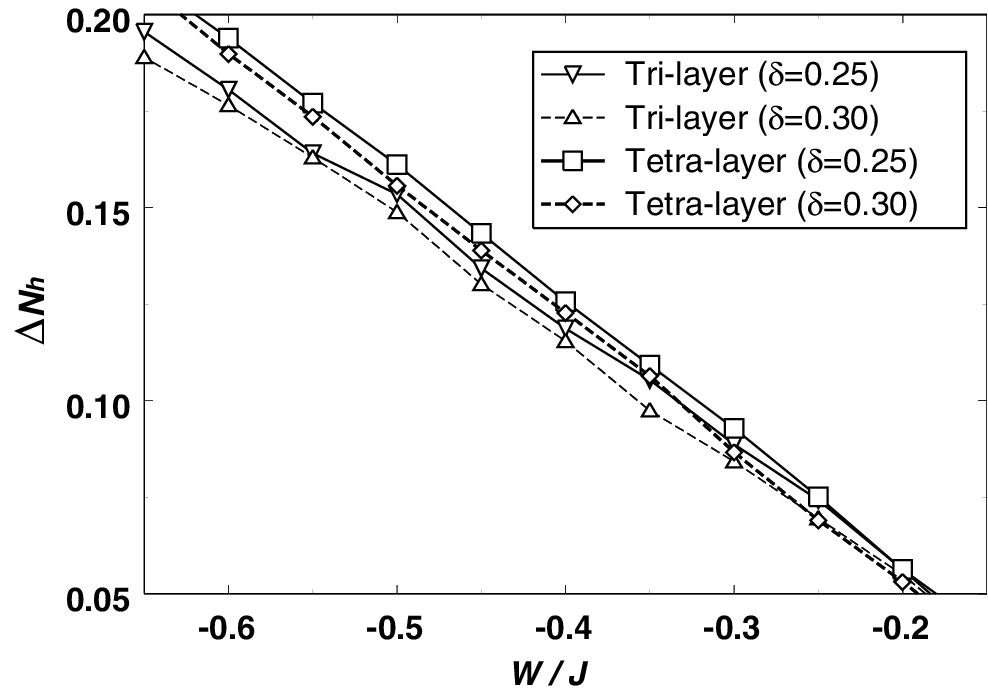}}
\vspace{4mm}
\caption{$W$ dependence of \ndif in the URVB state for the trilayer and the tetralayer with $\delta=0.25$ and 0.3. 
The values of the parameters are $J=0.14$ eV, $t/J=2.5$, $t'/J=-0.85$, $t''/J=0.575$, and $t_{\perp}/J=1.0$.
The averaged doping rate $\delta$ is defined as $\delta=\delta_{total}/n$, in which $\delta_{total}$ is a total hole density in a unit cell. 
}
\label{num}
\end{figure}

Before proceeding to the study of ARPES spectrum, it is necessary to determine the value of $W$ that is still a free parameter. 
We calculate \ndif as a function of $W$ in the URVB state of tri- and tetralayer systems with $\delta=$0.25 and 0.3. 
The results are shown in Fig.~\ref{num}. 
We determine a value of $W$ by comparing Fig.~\ref{num} to the results obtained by Kotegawa \etal\cite{Kotegawa} in NMR measurement. Their results of \ndif in the tri- and tetralayer samples are fitted by the following equations: 
\beqa
\mbox{\rm trilayer,}
&\hspace{12pt}&
\Delta N_h =-0.130+0.759\, \delta;\label{nmr3}\\
\mbox{\rm tetralayer,}
&\hspace{12pt}&
\Delta N_h=-0.150+0.934\, \delta.\label{nmr4}
\eeqa
By using Eqs.~(\ref{nmr3}) and (\ref{nmr4}), the values of $W$ that reproduce the NMR results are determined as
\beqa
&&\mbox{\rm trilayer},\nonumber\\ 
&&\Big\{
\begin{array}{r}
\delta=0.25 \rightarrow \Delta N_h= 0.06 \rightarrow W/J= -0.25,\\
\delta=0.30 \rightarrow \Delta N_h= 0.10 \rightarrow W/J= -0.35;\\
\end{array}\label{triW}\\
&&\mbox{\rm tetralayer},\nonumber\\
&&\Big\{
\begin{array}{r}
\delta=0.25 \rightarrow \Delta N_h= 0.08 \rightarrow W/J=-0.30,\\
\delta=0.30 \rightarrow \Delta N_h= 0.13 \rightarrow W/J=-0.45. 
\end{array}\label{tetraW}
\eeqa

\begin{figure}
\epsfxsize=7cm
\centerline{\epsffile{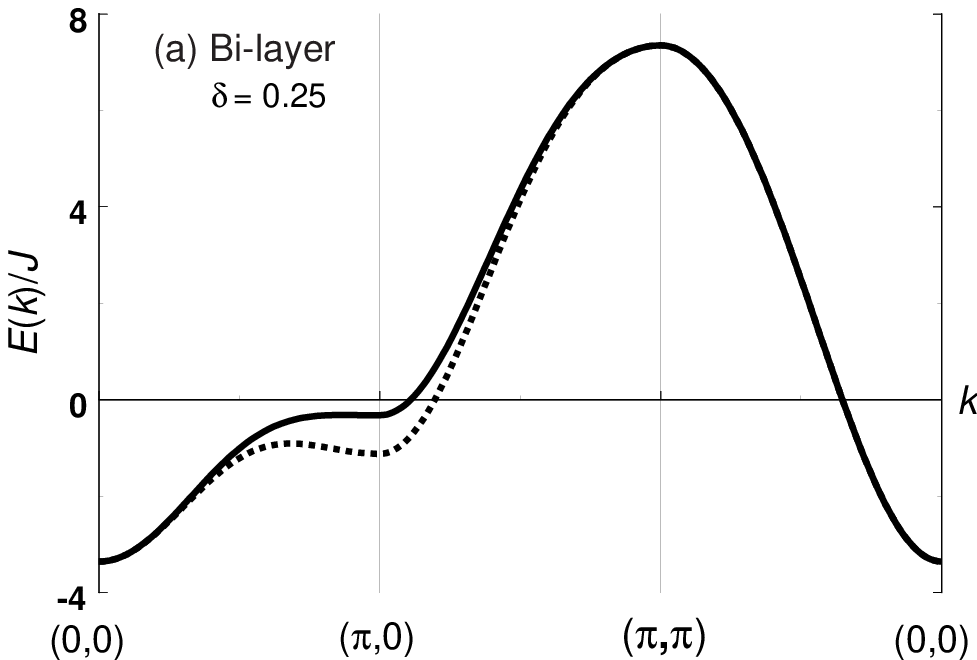}}
\vspace{4mm}
\epsfxsize=7cm
\centerline{\epsffile{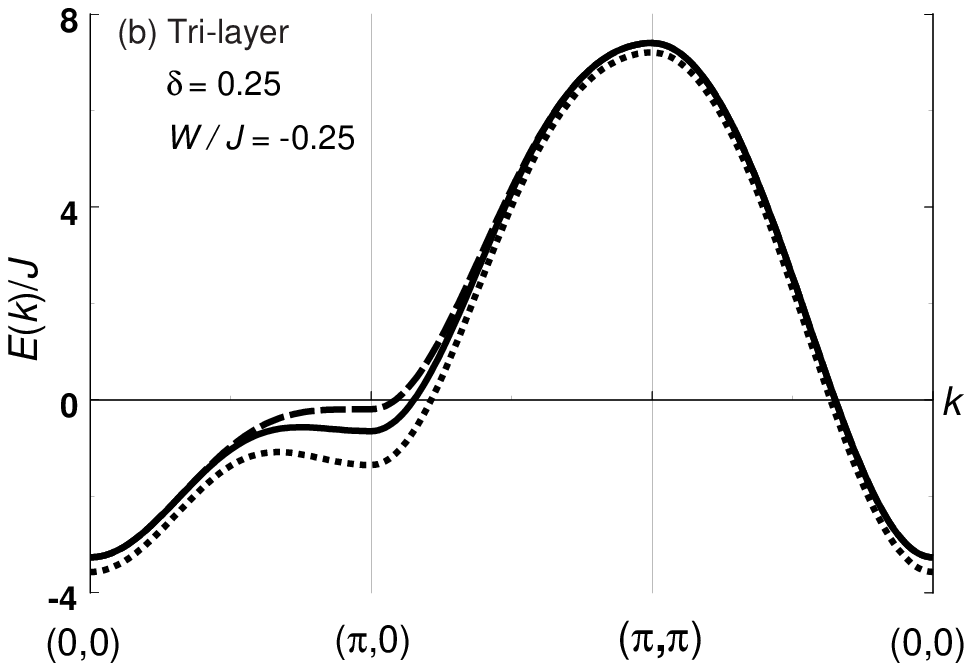}}
\vspace{4mm}
\epsfxsize=7cm
\centerline{\epsffile{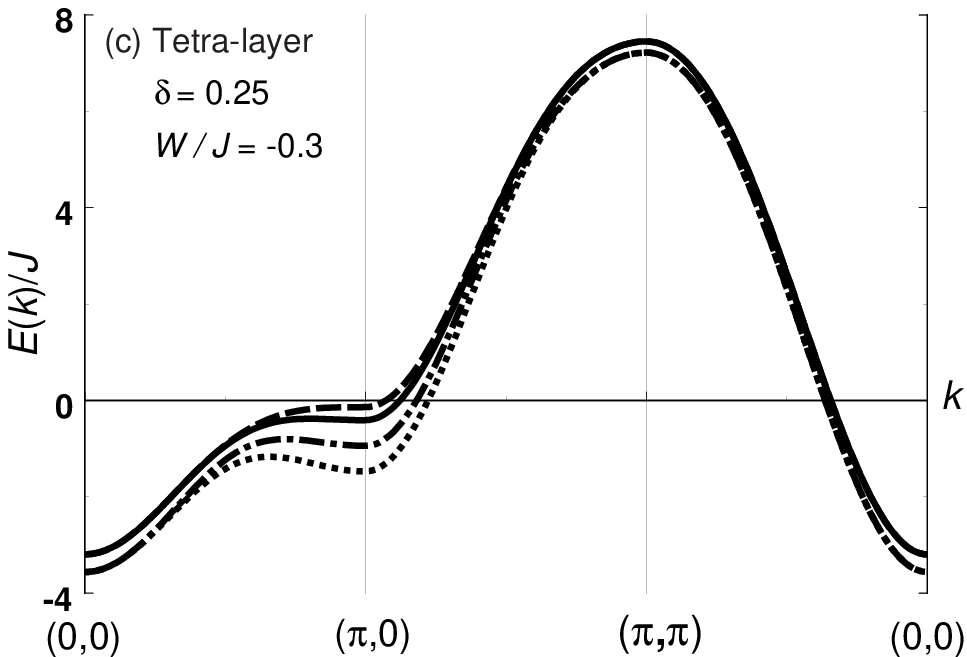}}
\vspace{4mm}
\caption{Dispersion relations of (a) bilayer, (b) trilayer, and (c) tetralayer systems for $\delta=0.25$ in the URVB state. 
The values of the parameters are, $J=0.14$ eV, $t/J=2.5$, $t'/J=-0.85$, $t''/J=0.575$, and $t_{\perp}/J=1.0$.
}
\label{disp}
\end{figure}

The dispersion relations of tri- and tetralayer systems are plotted together with the bilayer system ($W=0$) in Fig.~\ref{disp} for $\delta=0.25$.
Since the interlayer hopping is given by Eq.~(\ref{interhop}), the dispersion relations in the bilayer system are degenerate in the (0,0)-($\pi$,$\pi$) direction and the splitting becomes maximum at ($\pi$,0). 
The band with lower-binding energy indicated by the solid line in Fig.~\ref{disp}(a) is the AB band and the other one is the B band. 
The bilayer splitting at ($\pi$,0) for $\delta=0.25$ is 0.8$J$, which is given by $2 t_{\perp} g_t$. 
Most of experiments\cite{Feng1,Chuang1,Kordyuk,Chuang2,Feng3} and band calculations\cite{OKAnderson,Liechtenstein} show two holelike Fermi surfaces consistent with our dispersion relation. 
On the other hand, there are experimental\cite{Bogdanov} and theoretical studies\cite{Barnes} suggesting an electronlike Fermi surface, where the AB band crosses the Fermi level between (0,0) and ($\pi$,0). Such a dispersion relation would be achieved by small changes of parameters. 
In the trilayer system, there exist three bands.
Among them, one of the bands indicated by the dotted line in Fig.~\ref{disp}(b) is separated from the other two in the whole Brillouin zone due to the site potential. 
The middle band shown by the solid line in Fig.~\ref{disp}(b) is composed of only the OP's degrees of freedom as, $(|{\rm OP}_1\rangle - |{\rm OP}_2\rangle)/\sqrt{2}$, where $|{\rm OP}_1\rangle$ and $|{\rm OP}_2\rangle$ indicate the wave functions in the outer planes. 
We refer to this band as the $\lq\lq$antisymmetric (AOP) band$"$.
The wave functions of remaining two bands  are $f(W)(|{\rm OP}_1\rangle + |{\rm OP}_2\rangle)/\sqrt{2}\pm g(W)|{\rm IP}\rangle$, where $|{\rm IP}\rangle$ indicates the IP's wave function. 
The coefficients $f(W)$ and $g(W)$ are functions of $W$. 
The positive (negative) sign corresponding to the dotted (broken) line in Fig.~\ref{disp}(b) is referred to as the $\lq\lq$IP band$"$ ($\lq\lq$symmetric (SOP) band$"$).
In the tetralayer system, OP's and IP's combine to give symmetric and antisymmetric combinations as
$|{\rm OP}_{\pm}\rangle=(|{\rm OP}_1\rangle \pm |{\rm OP}_2\rangle)/\sqrt{2}$ and 
$|{\rm IP}_{\pm}\rangle=(|{\rm IP}_1\rangle \pm |{\rm IP}_2\rangle)/\sqrt{2}$. 
The four bands shown in Fig.~\ref{disp}(c) have the following combinations at ($\pi$,0) from the top to the bottom,
$|\alpha\rangle=f_1|{\rm IP}_{-}\rangle+g_1|{\rm OP}_{-}\rangle$, 
$|\beta \rangle=f_2|{\rm IP}_{+}\rangle-g_2|{\rm OP}_{+}\rangle$, 
$|\gamma\rangle=f_3|{\rm IP}_{-}\rangle-g_3|{\rm OP}_{-}\rangle$, and 
$|\delta\rangle=f_4|{\rm IP}_{+}\rangle+g_4|{\rm OP}_{+}\rangle$,
where $f_i$ and $g_i$ ($i=1\sim4$) are functions of $W$. 
\begin{figure}[t]
\epsfxsize=7.5cm
\centerline{\epsffile{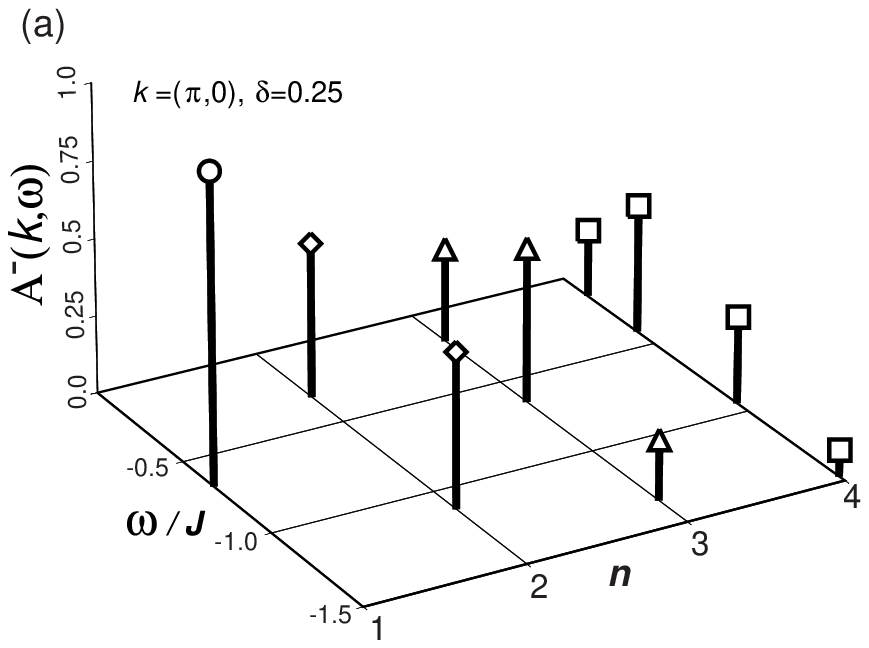}}
\vspace{4mm}
\epsfxsize=7.5cm
\centerline{\epsffile{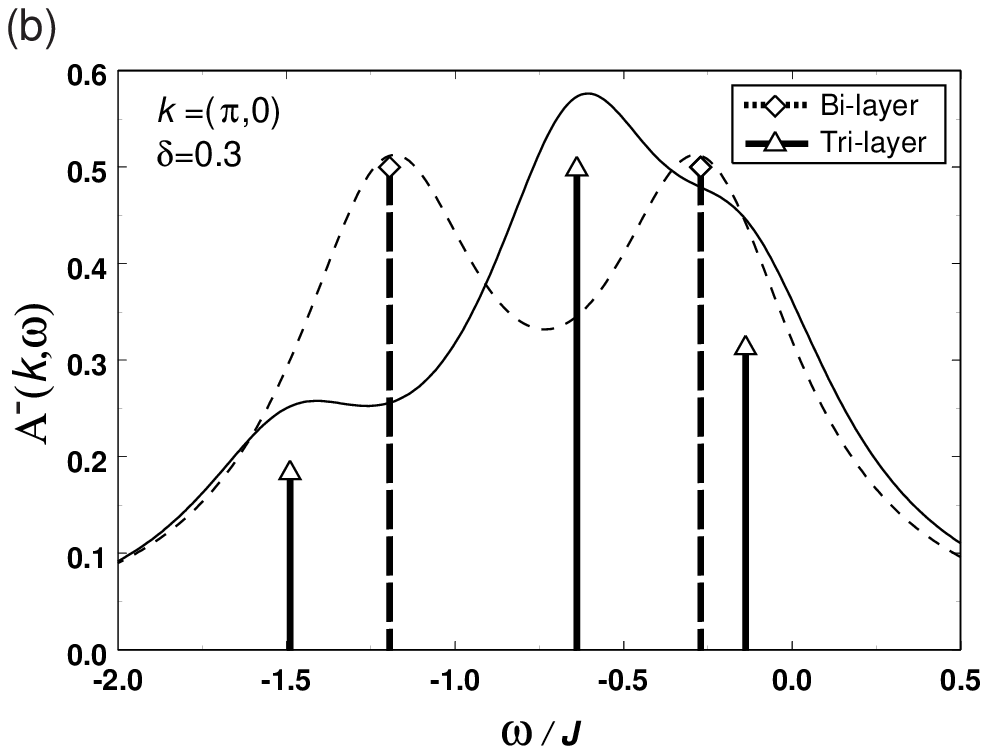}}
\vspace{4mm}
\caption{The $n$ dependence of the electron-removal spectral function in the normal phase for $J=0.14$ eV, $t/J=2.5$, $t'/J=-0.85$, $t''/J=0.575$, and $t_{\perp}/J=1.0$. 
The components in the \cuo plane nearest to the surface are plotted 
(a) on the $n$-$\omega$ plane for $\delta=$0.25, and 
(b)  as functions of $\omega$ for $\delta=$0.30 for the bi- and trilayer systems. The thin solid (broken) curve is obtained by performing a Lorentzian broadening with a width of 0.35$J$ for the $\delta$ functions in the bi- (tri-) layer system.  
In (a) and (b), the values of $W$ follow Eqs.~(\ref{triW}) and (\ref{tetraW}).}
\label{AkwN}
\end{figure}

In the following, we assume that the ARPES experiments in Bi compounds mainly detect the single-particle excitation in the \cuo plane next to the BiO layer. 
The ARPES measurement is sensitive to a surface of sample due to a short escape depth of outgoing electron, which is about 10 \AA$\;$ or less depending on its kinetic energy in the range of 20$-$200 eV.\cite{Shen-Dessau} 
The cleaved surface is considered to be the Bi-O layer, from which the distance to the IP of trilayer compounds is also about 10 \AA.\cite{Ihara,Parkin}  
Therefore the ARPES spectra in the tri- and tetralayer compounds should be dominated by the OP's contribution. 
Moreover, the spectra depend on the incident photon energy \hnu and the photoemission matrix element.\cite{Chuang2,Feng3,Dahnken,Lindroos} 
In the case of overdoped bilayer samples, an antibonding (AB) band has a sharp peak at ($\pi$,0) for \hnu$\sim$ 22 eV.
Although a bonding (B) band is unclear for \hnu$\sim$ 22 eV, 
it becomes clear for \hnu$\sim$ 32 eV (Ref. 23) and 47 eV.\cite{Chuang2}

The $n$ dependence of the OP contributions in the electron-removal spectral function $A^-(k,\omega)$, at ($\pi$,0) is plotted in Fig.~\ref{AkwN} for $t_{\perp}/J=1.0$. 
The energy is measured from the Fermi level. 
In the bilayer system, the two quasiparticle peaks clearly separate each other for both doping rates. 
The peak with lower-binding energy is the AB band and the other is the B band. 
In the trilayer system, the OP's combine to give symmetric and antisymmetric combinations, the latter of which is the AOP band as defined above and is located at $\omega/J$=0.64. Only the former can hybridize with the IP, and produces the SOP band ($\omega/J$=0.14) and the IP band ($\omega/J$=1.49).
The total splitting in the trilayer compound ($\Delta\omega/J$=1.35) is larger than in the bilayer compound ($\Delta\omega/J$=0.9), but the SOP-AOP splitting ($\Delta\omega/J$=0.5) is smaller than the bilayer splitting. Therefore the trilayer spectrum seems to be narrower than the bilayer spectrum as shown in Fig.~3(b), in which the appropriate broadening is given.

The contributions to the spectral weight from the OP and IP are listed in Table~\ref{tri-weight} for the three bands in the trilayer system together with their binding energies. 
The OP contribution distributes among the three bands as SOP:AOP:IP$\sim$3:5:2, while the IP contribution is obtained as SOP:AOP:IP$\sim$4:0:6. 
If the electron escape depth is longer than 10 \AA$~$, 
the weight of the IP band increases as a result of the IP contribution to the ARPES spectrum. This makes the spectrum broader than the bilayer spectrum as easily expected from the Fig~3(b). 
Thus, the broadness might be a measure of the escape depth. 
\begin{table}[tbp]
\caption{The binding energies and the ARPES spectral weights at $k=(\pi,0)$ in a trilayer compound for $\delta$=0.3. Each contribution from the IP and the OP is listed. }
\begin{tabular}{lccc}	
  & $\omega/J$  & OP$^a$ & IP$^b$\\
\hline
SOP-band  & 0.14 & 0.31 & 0.37 \\
AOP-band  & 0.64 & 0.50 & 0.00 \\
IP-band & 1.49 & 0.19 & 0.63 
\label{tri-weight}
\end{tabular}
{\footnotesize $a$ A \cuo layer nearest to the surface. 
  
$b$ A \cuo layer next nearest to the surface.}
\end{table}

Each spectral weight in the tetralayer compound is listed in Table.~\ref{tetra-weight}. 
The dominant three peaks in the OP, \ie, $\alpha$, $\beta$ and $\gamma$ band, may result in a broad peak due to the line overlap. 
Including the IP contribution, it may become more difficult to distinguish the peaks. 
\begin{table}[tbp]
\caption{The binding energies and the ARPES spectral weights at $k=(\pi,0)$ in a tetralayer compound for $\delta$=0.3. Each contribution from the IP and the OP is listed.}
\begin{tabular}{lccc}	
  & $\omega/J$  & OP$^a$ & IP$^b$\\
\hline 
$\alpha$-band & 0.08 & 0.25 & 0.25 \\
$\beta$-band  & 0.35 & 0.43 & 0.07 \\
$\gamma$-band  & 1.00 & 0.25 & 0.25 \\
$\delta$-band  & 1.65 & 0.07 & 0.43
\label{tetra-weight}
\end{tabular}
{\footnotesize $a$ A \cuo layer nearest to the surface.

$b$ A \cuo layer next nearest to the surface.}
\end{table}

\section{Superconducting Amplitudes}\label{super}
\begin{figure}
\epsfxsize=7.5cm
\centerline{\epsffile{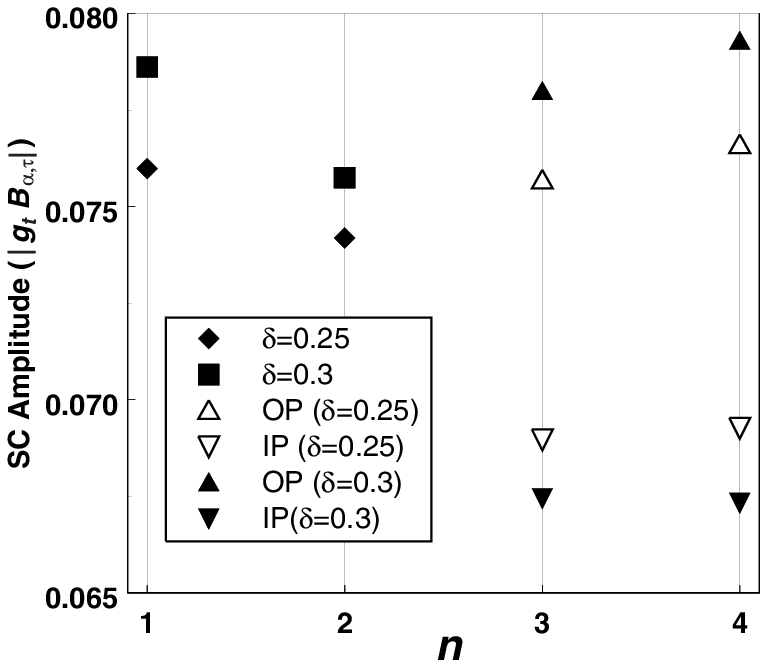}}
\vspace{4mm}
\caption{The $n$ dependence of the SC amplitude at $\delta=0.3$.
For $n\ge$ 3, the up and down triangle indicate the amplitude in the OP and the IP, respectively. At $n=$2, both layers have the same amplitude. 
The value of parameters are, $J=0.14$ eV, $t/J=2.5$, $t'/J=-0.85$, $t''/J=0.575$, and $t_{\perp}/J=1.0$. 
The values of $W$ follow Eqs.~(\ref{triW}) and (\ref{tetraW}).
}
\label{SCamp}
\end{figure}

In this section, we consider the superconducting (SC) state, where the SRVB state is included in addition to the URVB state. 
As observed in the tetralayer compounds by NMR experiment, the IP and the OP have different $T_0$'s, which are ascribed to the different hole concentration in each layer.\cite{Tokunaga} 
In the trilayer compounds, the different $T_0$'s are not still found,\cite{Kotegawa} although there is a finite \ndif, which is smaller than \ndif in the tetralayer compounds.
To understand the SC state in the multilayer cuprates, 
we calculate the SC amplitude in each layer defined as $|g_t B_{\alpha,\tau}|$, and show the $n$ dependence. 
In the following, we assume that all \cuo layers are in the SC phase, and the value of $W$ in the SC phase is the same as that in the normal phase. 
The results are shown in Fig.~\ref{SCamp}.
The SC gap has the $d_{x^2-y^2}$ symmetry in each layer for any $n$. 
The OP's in both the tri- and tetralayer systems have different superconducting amplitude from the IP, while each layer in the bilayer systems has the same amplitude. 
For $n\ge 3$, the difference of SC amplitude slightly increases with increasing $n$. 
Since the difference in the trilayer systems is comparable to that in the tetralayer systems, the different $T_0$ could be observed in experiment.

We note that, although the SC amplitude is larger in the OP than the IP as seen in Fig.~\ref{SCamp}, the relative magnitude depends on the intralayer's parameters, which modify the phase diagram of ground state in the monolayer system.\cite{Tanamoto} 
In the present calculation, moreover, the site potential is fixed at the value in the normal phase. 
It is also possible that the site potential depends on the hole densities and is determined self-consistently. 
If such mechanisms are taken into account, the hole distribution among the layers may changes below \tc$\,$ and the valance of SC amplitudes can be modified.  
These possibilities are studied elsewhere.

We did not mention quasiparticle excitations in the SC phase.
The coherent SC peak in Bi2212 seems to split into two peaks.\cite{Feng1}  
There are some interpretations of the bilayer splitting in the SC phase.\cite{Feng1,Barnes} 
To clarify its origin, more experimental studies on the multilayer cuprates are needed.

\section{Summary}\label{summary}
We have studied the electronic states and the superconductivity in multilayer cuprates by use of the resonating valence bond wave function and the Gutzwiller approximation for the two-dimensional multilayer \tj model. 
In this model, the one-particle hopping connects the chemically different \cuo planes in the tri- and tetralayer systems. 
We calculated the electron-removal spectra in the \cuo plane next to the BiO layer, since the ARPES is sensitive to the surface of sample.
We find that the ARPES spectra in the trilayer systems at ($\pi$,0) should be narrower than that in the bilayer systems, and wider than that in the monolayer systems.
This is consistent with recent experimental data.\cite{Feng4}

The superconducting (SC) amplitude of each layer is shown as a function of $n$. 
In bilayer system, both layers have the same SC amplitude. 
In tri- and tetralayer systems, the SC amplitude of the OP is different from that in the IP due to the different hole concentration. 
Such a difference could be observed as two different $T_0$ also in the trilayer compounds. 

\acknowledgments
The authors would like to thank Z.-X. Shen and D. L. Feng for valuable discussions. This work was supported by the Grant-in-Aid from Ministry of Education, Culture, Sports, Science and Technology of Japan, and CREST. One of authors (S.M.) acknowledges support of the Humboldt Foundation.
\appendix
\section*{Gutzwiller Approximation and RVB Wave Function in Multilayer Systems}\label{GWA}
The RVB-mean-field theory with the Gutzwiller approximation showed that the $d_{x^2-y^2}$ wave superconductivity can be stabilized in the two-dimensional $t$-$J$ model.\cite{FCZhang}
The Gutzwiller approximation has been recently generalized to the coexistent state of the $d$-wave SC and AF order.\cite{Ogata} 
In this section, we summarize the process to treat the constraint within this approximation.

Explicitly describing the constraint, 
the Hamiltonian Eq.~(\ref{Htot})  has the following relation to the nonconstrained Hamiltonian:
\beq
\hat{H}=P_G H P_G,
\eeq
where $P_G$ is the Gutzwiller's projection operator and defined as
\beq
P_G=\prod_i (1-\hat{n}_{i\uparrow}\hat{n}_{i\downarrow}).
\eeq
We assume the projected RVB wave function 
\beq
\label{vwf}
 |\psi\rangle
 	=P_G|\psi_0(\chi_{\alpha,\tau},B_{\alpha,\tau},\mu)\rangle,
\eeq
where $\chi_{\alpha,\tau}$ and $B_{\alpha,\tau}$ 
are the variational parameters relating to URVB and SRVB, respectively, and 
$|\psi_0(\chi_{\alpha,\tau},B_{\alpha,\tau},\mu)\rangle$ 
is a mean-field wave function with URVB and SRVB orders.
Using these notations, the variational energy 
$E_{\rm var} = \langle {\widehat H}\rangle$ 
is rewritten as 
$E_{\rm var} = \langle H \rangle_0$,
where the parameters $t$ and $J$ in $\widehat{H}$ are replaced with 
\beqa
&&
\hat{t}= g_t\cdot t, \;
\hat{t}'= g_t\cdot t', \;
\hat{t}''= g_t \cdot t'', \;
\hat{t}_{\perp}= g_t \cdot t_{\perp},\nonumber\\
&&
\hat{J}= g_J \cdot J,
\eeqa
and
\beq
g_t = 2\delta/(1+\delta), \;g_J=4/(1+\delta)^2.
\eeq
In the Gutzwiller approximation, 
the effects of the projection are statistically averaged and renormalized into the expectation values as follows:
\beq
 \langle c_{i\sigma}^{\dagger} c_{j\sigma}\rangle =
 g_t\langle c_{i\sigma}^{\dagger} c_{j\sigma}\rangle_{0}, \quad
 \langle {\vec{S}}_i\cdot{\vec{S}}_j\rangle =
 g_J\langle {\vec{S}}_i\cdot {\vec{S}}_j\rangle_{0},  
\label{defgs}
\eeq
where $\langle \cdots \rangle_0$ represents the expectation value 
in terms of 
$|\psi_0\rangle = 
|\psi_0(\chi_{\alpha,\tau},B_{\alpha,\tau},\mu)\rangle$,
and $\langle \cdots \rangle$ represents the normalized expectation 
value in $|\psi \rangle = P_G|\psi\rangle_0$; 
\beq
\langle \widehat{\cal O} \rangle \equiv 
\frac{\langle \psi | \widehat{\cal O} | \psi\rangle }
{\langle \psi| \psi \rangle } 
=\frac{\langle \psi_0 | P_G \widehat{\cal O} P_G  | \psi_0\rangle }
{\langle \psi_0 | P_G P_G  | \psi_0\rangle } .
\eeq
By considering the renormalization factors of coupling constant, the remaining work is to solve the self-consistent equations, in which the renormalization factors are also determined self-consistently. 
In the multilayer systems, the constraint is included in the interlayer hopping and renormalizes $t_{\perp}$.
Eder \etal numerically showed that the $t_{\perp}$ is renormalized by an in-plane quasiparticle weight.\cite{Eder} 
Such an effect is reproduced by the Gutzwiller factor $g_t$.

Here, we summarize the eigenvalue problem in the SC state of multilayer systems.\cite{Blaizot}
Neglecting the details, the Hamiltonian generally has the following form: 
\beqa
H
	&=&
		 \sum_{\alpha,\beta,k,\sigma} 
		 	A \; c_{\alpha,k,\sigma}^{\dag}c_{\beta,k,\sigma}\nonumber\\
		&&+\sum_{\alpha,k} 
			\left(B^*\:
				c_{\alpha,-k,\downarrow}c_{\alpha,k,\uparrow}
	             +B  \;
	             c_{\alpha, k,\uparrow}^{\dag}c^{\dag}_{\alpha,-k,\downarrow}
			      \right),\nonumber\\
	&=&
		\sum_k
			\phi^{\dag}
			\left(
				\begin{array}{cc}
				A  & B  \\
				B^*&-A^*
				\end{array}
			\right)
			\phi
		+{\rm Tr} A,\nonumber\\
	&\equiv&
		\sum_k
			\phi^{\dag} M \phi +{\rm Tr}A,		\label{genHamil}\\
\phi^{\dag}
	&\equiv&
		\left(c_{\alpha,k,\uparrow}^{\dag},c_{\alpha,-k,\downarrow}\right),
\eeqa
where
\beq
A^{\dag}=A, \;\; ^t B=B.
\eeq
The matrix $M$ is hermite and satisfies the following relations:
\beqa
\sigma_3\sigma_1 M \sigma_1\sigma_3 
&=& 
-M^*, \label{Mident2}\\
M^{\dag}&=&M.
\eeqa
We obtain the eigenvalue and the eigenvectors by solving 
\beq
MV_n=\omega_n V_n
\equiv
\omega_n
\left(
\begin{array}{c}
X_n\\
Y_n
\end{array}
\right).  \label{Egeq3}
\eeq
By taking account of Eq.~(\ref{Mident2}), the complex conjugate of Eq.~(\ref{Egeq3}) satisfies the following equation: 
\beq
M\sigma_1\sigma_3V^*_n=-\omega_n \sigma_1\sigma_3V^*_n.\label{Egeq4}
\eeq
Therefore $W_n\equiv\sigma_1\sigma_3 V^*_n$ is also the eigenvector of $M$.
If $V_n$ belongs to a positive eigenvalue $\omega_n > 0 $, another eigenvector $W_n$ belongs to a negative eigenvalue $-\omega_n$. 
By use of $b^{\dag}_n$ and $b_n$ defined below:
\beqa
\left(
	\begin{array}{c}
	b_n\\
	^t b^{\dag}_n
	\end{array}
\right)
	&=& \sum
	\left(
		\begin{array}{cc}
		X^{*}_{n,\alpha}&Y^*_{n,\alpha}\\
	   -Y    _{n,\alpha}&X  _{n,\alpha}
		\end{array}
	\right)
	\left(
	\begin{array}{c}
		c_{\alpha,\uparrow}\\
		^t c^{\dag}_{\alpha,\downarrow}
	\end{array}
	\right),
\label{Bogoliubov}
\eeqa
the Hamiltonian Eq.~(\ref{genHamil}) is diagonalized as
\beq
H
	=
	 2 \sum_{n>0} \omega_n b^{\dag}_n b_n
	  -\sum_{n>0} \omega_n
	  +{\rm Tr} A.							\label{digHmail}
\eeq
By using Eq.~(\ref{Bogoliubov}), the electron operator is 
\beq
c_{\alpha,\uparrow}
	=
	\sum_n
	\left(
	X_{\alpha,n}b_n-Y^*_{\alpha,n}\;^t b^{\dag}_n
	\right).
\eeq
The order parameters and the spectral weight are calculated as
\beqa
\langle 0| c^{\dag}_{\alpha,\sigma}c_{\beta,\sigma} |0 \rangle 
&=&
\sum_n   Y^*_{\beta,n} Y_{n,\alpha},\\
\langle 0| c^{\dag}_{\alpha,\sigma}c^{\dag}_{\beta,-\sigma} |0 \rangle 
&=&
\sum_n - X^*_{\beta,n} Y_{n,\alpha},\\
| \langle n| c_{\alpha,\sigma} |0 \rangle |^2
&=&
 | Y_{\alpha,n}|^2.
\eeqa


\begin{references}
\bibitem{Ihara}
H. Ihara, R. Sugise, M. Hirabayashi, N. Terada, M. Jo, K. Hayashi, A. Negishi, M. Tokumoto, Y. Kumira, and T. Shimomura, 
\jnl{Nature (London)}{334}{510}{1988}

\bibitem{Parkin}
S. S. P. Parkin, V. Y. Lee, A. I. Nazzal, R. Savoy, and R. Beyers,
\jnl{\PRL}{61}{750}{1988}

\bibitem{Kondo}
J. Kondo,
\jnl{\JPSJ}{58}{2884}{1989}

\bibitem{Ohta2}
Y. Ohta, T. Tohyama, and S. Maekawa,
\jnl{\PRB}{43}{2968}{1991}

\bibitem{Ohta}
Y. Ohta and S. Maekawa,
\jnl{\PRB}{41}{6524}{1990}

\bibitem{DiStatio}
M. Di Stasio, K. A. M${\rm \ddot
{u}}$ller, and L. Pietronero,
\jnl{\PRL}{64}{2827}{1990}

\bibitem{Haines}
E. M. Haines and J. L. Tallon,
\jnl{\PRB}{45}{3172}{1992}

\bibitem{Trokiner}
A. Trokiner, L. Le Noc, J. Schneck, A. M. Pougnet, R. Mellet, J. Primot, H. Savary, Y. M. Gao, and S. Aubry,
\jnl{\PRB}{44}{2426}{1991}

\bibitem{Statt}
B. W. Statt and L. M. Song,
\jnl{\PRB}{48}{3536}{1993}

\bibitem{Magishi}
K. Magishi, Y. Kitaoka, G. Aheng, K. Asayama, K. Tokiwa, A. Iyo, and H. Ihara,
\jnl{\JPSJ}{64}{4561}{1995}

\bibitem{Julien}
M.-H. Julien, P. Carretta, M. Horvati${\rm \acute{c}}$, C. Berthier, Y. Berthier, P. S${\rm \acute{e}}$granan,
 A. Carrington, and D. Colson,
\jnl{\PRL}{76}{4238}{1996} 

\bibitem{Kotegawa}
H. Kotegawa, Y. Tokunaga, K. Ishida, G.-q. Zheng, Y. Kitaoka, H. Kito, A. Iyo, K. Tokiwa, T. Watanabe, and H. Ihara,
\jnl{\PRB}{64}{064515}{2001}

\bibitem{Zheng}
G.-q. Zheng, Y. Kitaoka, K. Asayama, K. Hamada, H. Yamauchi, and S. Tanaka,
\jnl{\JPSJ}{64}{3184}{1995}

\bibitem{Chakravarty}
S. Chakravarty, A. Sudbo, P. W. Anderson, and S. Strong,
\jnl{Science}{261}{337}{1993}

\bibitem{OKAnderson}
O. K. Andersen, A. I. Liechtenstein, O. Jepsen, and F. Paulsen, 
\jnl{\JPCS}{56}{1573}{1995}

\bibitem{Liechtenstein}
A. I. Liechtenstein, O. Gunnarsson, O. K. Andersen, and R. M. Martin,
\jnl{\PRB}{54}{12505}{1996} 
\bibitem{Feng1}
D. L. Feng, N. P. Armitage, D. H. Lu, A. Damascelli, J. P. Hu, P. Bogdanov, A. Lanzara, F. Ronning, K. M. Shen, H. Eisaki, C. Kim, Z.-X. Shen, J.-i. Shimoyam, and K. Kishio
\jnl{\PRL}{86}{5550}{2001}

\bibitem{Chuang1}
Y. -D. Chuang, A. D. Gromko, A. Fedorov, D. S. Dessau, Y. Aiura, K. Oka, Y. Ando, H. Eisaki, and S. I. Uchida,
\jnl{\PRL}{87}{117002}{2001}

\bibitem{Kordyuk}
A. A. Kordyuk, S. V. Borisenko, M. S. Golden, S. Legner, K. A. Nenkov, M. Knupfer, J. Fink, H. Berger, R. Follath, and L. Forro,
\jnl{\PRB}{66}{014502}{2002}

\bibitem{Feng2}
D. L. Feng, A. Damascelli, K. M. Shen, N. Motoyama, D. H. Lu, H. Eisaki, 
K. Shimizu, J. -i. Shimoyama, K. Kishio, N. Kaneko, M. Greven, G. D. Gu, 
X. J. Zhou, C. Kim, F. Ronning, N. P. Armitage, and, Z.-X. Shen,  
\jnl{\PRL}{88}{107001}{2002}

\bibitem{Sato}
T. Sato, H. Matsui, S. Nishina, T. Takahashi, T. Fujii, T. Watanabe, and A. Matsuda,
cond-mat/0108415.

\bibitem{Chuang2}
Y. -D. Chuang, A. D. Gromko, A. V. Fedorov, Y. Aiura, K. Oka, Yoichi Ando, and D. S. Dessau, 
cond-mat/0107002.

\bibitem{Feng3}
D. L. Feng, C. Kim, H. Eisaki, D. H. Lu, K. M. Shen, F. Ronning, N. P. Armitage, A. Damascelli, N. Kaneko, M. Greven, J. -i. Shimoyama,  K. Kishio,  R. Yoshizaki, G. D. Gu, and Z.-X. Shen,
\jnl{\PRB}{65}{220501}{2002}

\bibitem{Dahnken}
C. Dahnken and R. Eder,
cond-mat/0109036.

\bibitem{Lindroos}
M. Lindroos, S. Sahrakorpi, and A. Bansil,
\jnl{\PRB}{65}{054514}{2002}

\bibitem{Tokunaga}
Y. Tokunaga, K. Ishida, Y. Kitaoka, K. Asayama, K. Tokiwa, A. Iyo, and H. Ihara,
\jnl{\PRB}{61}{9707}{2000}

\bibitem{Suhl}
H. Suhl, B. T. Matthias, and L. R. Walker,
\jnl{\PRL}{3}{552}{1959}

\bibitem{Kim}
C. Kim, P. J. White, Z.-X. Shen, T. Tohyama, Y. Shibata, S. Maekawa, B. O. Wells, Y. J. Kim, R. J. Birgeneau, and M. A. Kastner,
\jnl{\PRL}{80}{4245}{1998}

\bibitem{Tohyama1}
T. Tohyama and S. Maekawa,
Supercond. Sci. Technol. {\bf 13}, R17 (2000).

\bibitem{Shen}
Z.-X. Shen and J. R. Schrieffer,
\jnl{\PRL}{78}{1771}{1997}

\bibitem{Campuzano}
J. C. Campuzano, H. Ding, M. R. Norman, H. M. Fretwell, M. Randeria, A. Kaminsk
\jnl{\PRL}{83}{3709}{1999}

\bibitem{Bogdanov}
P. V. Bogdanov, A. Lanzara, X. J. Zhou, S. A. Kellar, D. L. Feng, E. D. Lu, 
H. Eisaki, J. -I. Shimoyama, K. Kishi, Z. Hussain, and Z.-X. Shen, 
\jnl{\PRB}{64}{180505}{2001}

\bibitem{Barnes}
S. Barnes and S. Maekawa,
cond-mat/0111205.

\bibitem{Shen-Dessau}
Z.-X. Shen and D. S. Dessau,
\jnl{Phys. Rep.}{253}{1}{1995}

\bibitem{Tanamoto}
T. Tanamoto, H. Kohno, and H. Fukuyama,
\jnl{\JPSJ}{63}{2739}{1994}

\bibitem{Feng4}
D. L. Feng and Z.-X. Shen,
(private communication).

\bibitem{FCZhang}
F. C. Zhang, C. Gros, T. M. Rice, and H. Shiba,
Supercond. Sci. Technol. {\bf 1}, 36 (1996).

\bibitem{Ogata}
M. Ogata and A. Himeta,
cond-mat/0003465.

\bibitem{Eder}
R. Eder, Y. Ohta, and S. Maekawa,
\jnl{\PRB}{51}{3265}{1995}

\bibitem{Blaizot}
J.-P. Blaizot and G. Ripka,
{\it Quantum Theory of Finite Systems} 
(MIT Press, Cambridge, MA, 1986). 
\end{references}
\end{document}